\documentclass[review,10pt]{elsarticle}

  \usepackage{graphicx}

\usepackage{amssymb}

\biboptions{sort&compress}
\journal{Journal of Crystal Growth}

\begin{document}

\begin{frontmatter}



\title{Laser-heated pedestal growth of cerium doped calcium scandate crystal fibers}



\author{J. Philippen}
\ead{jan.philippen@ikz-berlin.de}
\corref{cor1}
\author{C. Guguschev}
\author{R. Bertram}
\author{D. Klimm}

\cortext[cor1]{Corresponding author}

\address{Leibniz Institute for Crystal Growth, Max-Born-Str. 2, 12489 Berlin, Germany}

\begin{abstract}

Ce$^{3+}$ doped oxide materials are promising for optical emission in the green spectral range. The growth of CaSc$_{2}$O$_{4}$:Ce$^{3+}$ single crystals is reported here for the first time. Laser heated pedestal growth (LHPG) proved to be suitable for this refractive material, if performed in nitrogen of 99.999\% purity. If the oxygen content of the growth atmosphere is substantially larger, Ce$^{4+}$ is formed, which shows no useful optical emission. If the oxygen content is substantially lower, severe evaporation of calcium impedes stable crystal growth. Thermodynamic equilibrium calculations allowed to describe evaporation of species and cerium dopant charging under different growth conditions. The evaporation could be investigated by quadrupole mass spectrometry of emanating gases and by chemical analysis of fibers with ICP-OES. The congruent melting point was confirmed by DTA at 2110 degrees centigrade. Photoluminescence spectrometry of fibers revealed the dependence of optical emission in the green spectral range on growth conditions.

\end{abstract}

\begin{keyword}


A2. Laser heated pedestal growth \sep B1. Oxides \sep B2. Phosphors \sep B3. Solid state lasers

\end{keyword}

\end{frontmatter}


\section{Introduction}
\label{intro}

Among a vast variety of oxides that are used for light emission \cite{DORENBOS2001}, Ce$^{3+}$ doped materials are especially interesting because this ion shows broad green luminescence with a peak at 515\,nm \cite{SHIMOMURA2007}, if it is incorporated e.g. into CaSc$_{2}$O$_{4}$. This emission, just in the middle of the visible spectral region, makes Ce$^{3+}$ doped solid state lasers interesting for technical applications. According to Dorenbos \cite{DORENBOS2000}, the excitation and emission in a wide wavelength region is either due to a great crystal field depression or due to large Stokes shift. The luminescence properties of cerium doped CaSc$_{2}$O$_{4}$ can be compared to the isostructural SrY$_{2}$O$_{4}$ \cite{MANIVANNAN2003} (CaFe$_{2}$O$_{4}$ structure type, $Pnam$). The crystal structure of calcium scandate, investigated by Carter and Feigelson \cite{CARTER1964}, can be described as a type of AM$_{2}$O$_{4}$ oxometallate showing a tunnel structure \cite{MUELLERBUSCHBAUM1963,MUELLERBUSCHBAUM2003}. According to Fechner \cite{Physik2011}, CaSc$_{2}$O$_{4}$:Ce$^{3+}$ could be a promising material for laser emission in the green spectral range. So far, studies on CaSc$_{2}$O$_{4}$:Ce$^{3+}$ were performed only with ceramic samples, because single crystals were not available.

For the incorporation of Ce$^{3+}$ into a ceramic light phosphor the material can be co-doped with alkali metal ions, such as Na$^{+}$ for charge compensation (Na$^{+}$,Ce$^{3+}$ $\leftrightarrow$ 2~Ca$^{2+}$). Unfortunately, this method cannot be established in crystal growth techniques from the melt, because alkali metals evaporate at the high melting temperature. Without this charge compensation it is questionable, whether Ce$^{3+}$ can be incorporated. Charge compensation through the formation of oxygen interstitials or calcium vacancies could be an alternative mechanism.

As the melting point of CaSc$_{2}$O$_{4}$ is about $2110^{\,\circ}$C \cite{GETMAN2000}, only a few crystal growth methods qualify for synthesis. The laser-heated pedestal growth (LHPG), based on the work of Burrus and Stone \cite{BURRUS1975}, is a suitable method for this material, because the independence of a crucible enables high temperature growth. Besides, atmospheres of different chemical nature (inert, oxidizing or reducing) can be established. A major contribution to modern laser-heated crystal growth was made by Fejer et al. and Feigelson \cite{FEJER1984, FEIGELSON1986}. In the last decades, LHPG became a powerful tool in new and conventional materials research \cite{Dhanarajet.Al.2010}.

Due to the localized heating by the laser beam, very high thermal gradients up to $10^{4}$\,K/cm are generated at the interface of the melt zone \cite{Uda1992}. According to this, rapid growth rates in the range of several mm/min can be established \cite{Dhanarajet.Al.2010}. High growth rates are preferred, when one or more components of the material are evaporating from the melt. As LHPG growth is a zone leveling process, the dopant concentration within the fiber is affected by evaporation of components and segregation \cite{Pfann1958}. With equal translation rates for the pedestal and the growing fiber and if evaporation can be neglected, the Pfann relation describes the segregation of a component in the SCF (single crystal fiber) \cite{Pfann1958}:
\begin{equation}
C_{s}(z)=C_{0} \cdot \left[1-(1-k_\mathrm{eff}) \, \exp\left(-\frac{k_\mathrm{eff} \cdot z}{l}\right)\right] \label{eq:k}
\end{equation}
with $C_{s}(z)$ and $C_{0}$ as concentration of the dopant in the growing fiber and the feed rod, $z$ the fiber axis, $l$ the zone length and $k_\mathrm{eff}$ the effective distribution coefficient. If there is no evaporation of components, the melt zone will be enriched with or depleted of the dopant for $k_\mathrm{eff}<1$ or $k_\mathrm{eff}>1$, respectively, until $C_{s}(z)$ equals $C_{0}$ (steady state). If there is evaporation of the dopant from the melt zone, $C_{s}(z)<C_{0}$ will hold always (case $k_\mathrm{eff}<1$). However, $C_{s}(z)$ will reach a constant value in steady state. In this study, an approach to the single crystal fiber (SCF) growth of cerium doped calcium scandate is given. 

\section{Experimental}
\label{exp}

\subsection{Materials and preparation}
\label{materials}

Samples were prepared from CaCO$_3$, Sc$_2$O$_3$ and CeO$_2$ powders with 99.99\% purity. Stoichiometric compositions of 0.01 CeO$_2$, 0.99 CaCO$_3$, 1.00 Sc$_2$O$_3$ for cerium doped samples and 1.00 CaCO$_3$, 1.00 Sc$_2$O$_3$ for cerium-free samples were calcinated at $900^{\,\circ}$C for 8 hours. 

The products were multiply ground and sintered at $1600^{\,\circ}$C to ensure complete reaction to calcium scandate. Phase purity was checked by XRD. The dried powders were pressed isostatically at 2000\,bar in a die, sintered at $1600^{\,\circ}$C for 24\,h and cut into rectangular prisms. These prisms of polycrystalline and monocrystalline calcium scandate were used as feed rods for LHPG.

\subsection{Crystal growth}
\label{crystal}

\begin{table}[htb]
\centering
\begin{tabular}{ll}
\hline
Parameter                           & LHPG \\
\hline
Growth rate                         & 1--2\,mm/min\\
Fiber diameter                      & 0.8--1.6\,mm \\
Fiber length                        & $\leq50$\,mm \\
Aspect ratio $h/d_\mathrm{cryst}$   & 1--3 \\
\hline \hline
Atmosphere                          & composition (mbar) \\
\hline
Oxidizing I                         & Ar (800), O$_2$ (200) \\
Oxidizing II                        & Ar (980), O$_2$ (20) \\
Inert                               & Ar (1000) \\
Reducing I$^*$                      & N$_2$ (1000) \\
Reducing II                         & N$_2$ (990), H$_2$ (10) \\
Reducing III                        & N$_2$ (950), H$_2$ (50) \\ 
\hline
\end{tabular}
\caption{Growth parameters and composition of atmospheres during LHPG process. All gases had 5N (99.999\%) nominal purity. If one assumes the rest impurity as air, an oxygen partial pressure of $p_{\mathrm{O}_2}\approx2\times10^{-6}$\,bar can be estimated for all atmospheres not already having O$_2$ as one major component \cite{Klimm09}. \newline $^*$It will be shown later (Fig.~\ref{fig:atmo}) that N$_2$ is slightly reducing, compared to Ar.}
\label{tab:growth_parameters}
\end{table}

The LHPG furnace in this study is similar to the furnace described by Fejer et al. \cite{FEJER1984}. SCFs were grown in different atmospheres from reducing over inert to oxidizing (Tab.~\ref{tab:growth_parameters}). At the beginning of the LHPG process, the top of the feed rod was melted to a small droplet. With low translation rate the seed crystal was directed to the droplet and contacted the melt. After seeding, both the seed rod and the feed rod have been moved upwards with similar translation rates until the fiber crystallized with a suitable diameter. With an appropriate ratio of translation rates the fiber crystallized with constant diameter. At the end of the process the seed rod was stopped and the fiber was pulled out from the melt. The power of the laser as well as the translation rate of the pedestal were adjusted to assure an aspect ratio $h/d_\mathrm{cryst}$ between 1 and 3 ($h$: height of the melt zone, $d_\mathrm{cryst}$: diameter of the SCF). The melt zone was overheated approximately 100 to 150\,K above the melting point. The temperature was measured with a pyrometer (accuracy 0.5\% of the determined surface temerature). The edge length of the pedestal base ranged from 0.8\,mm to 2.3\,mm. An overview of growth parameters and setup is given in Tab.~\ref{tab:growth_parameters}.

\subsection{Characterization}

For phase analysis an X-ray powder diffraction system (GE) with Bragg-Brentano geometry was used. The melting point was determined with a high-temperature DTA (Netzsch) in a lidded tungsten crucible in helium atmosphere. For elemental analysis we used ICP-OES (Spectro). 

\begin{figure}[htb]
\centering
\includegraphics[width=0.6\textwidth]{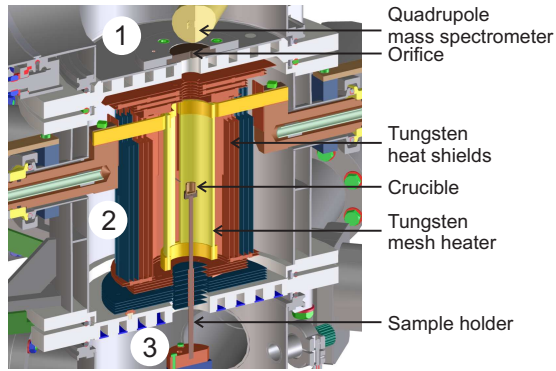} 
\caption{CAD model of the tungsten based MF-HTMS system. The numbers show different chambers inside the recipient with operating pressures at $10^{-7}$\,mbar (chamber~1) and $10^{-6}$\,mbar (chambers 2 \& 3).} 
\label{fig:christo}
\end{figure}

The evaporation of components was ex situ investigated with a home made multifunctional high-temperature mass spectrometer (MF-HTMS), described in detail elsewhere \cite{GUGUSCHEV2012}. Fig.~\ref{fig:christo} shows a CAD model of this system that consists of three separated chambers. Two turbo molecular pumps create a chamber pressure from $10^{-7}$\,mbar (chamber~1) to $10^{-6}$\,mbar (chambers~2 and 3). The sample container is symmetrically surrounded by a resistivity tungsten mesh heater. Tungsten and molybdenum heat shields create thermal insulation (chamber~2). In chamber~1 that is connected to chamber~2 via a 2\,mm orifice a quadrupole mass spectrometer (Pfeiffer Vacuum) with a crossbeam ion source detects a representative fraction of evaporating species. Remaining amounts of these species in chamber~2 condense at water-cooled parts, react with hot tungsten or molybdenum or are pumped out of the system. Under high vacuum conditions ($10^{-6}$\,mbar) SCFs were heated up to $2150^{\,\circ}$C in an open tungsten crucible. The electron impact energy was set to 70\,eV and a SEM voltage of 1300\,V was used for signal amplification. The temperature was adjusted manually and was measured with an optical pyrometer.

Photoluminescence measurements were performed using a commercial laser fluorescence spectrometer system of LTB (Lasertechnik Berlin GmbH). It consisted of a nitrogen laser (337\,nm wavelength, 500\,ps puls width) as excitation source and a spectrometer. The system is equipped with a detector combining a gated microchannel plate and a photodiode line camera. 

\subsection{Thermodynamic calculations}

Usually, under the very high temperatures $T$ that were used during growth thermodynamic equilibrium is reached quickly, and one can expect that calculations of thermodynamic equilibrium are a useful tool for the description of the process. Factsage\texttrademark ~\cite{FactSage2012}, an integrated thermodynamic data bank system and Gibbs free energy minimization program, was used for this purpose. Predominance diagrams, showing the stability of CeO$_x$ phases as function of $T$ and $\log[f_{\mathrm{O}_2}]$ (Fig.~\ref{fig:atmo}), as well as numerical data on the fugacity of evaporating species under the conditions of crystal growth (Tab.~\ref{tab:atmosphere}) were calculated. 

\section{Results and discussion}
\label{res}

\subsection{Crystal growth}
\label{growth}

The crystal diameter for LHPG processes is limited due to the large nonlinear temperature gradients in the furnace, typically exceeding 1000\,K/cm that induce thermal stress. It might be possible to increase the diameter by installing an afterheater or by enlargement of the laser beam size, but such work was outside the scope of this study. For the SCF growth itself, an almost inert (Ar, N$_2$) atmosphere proved to be beneficial. Under such conditions, the process was stable, reached quickly steady state, and evaporation losses were small. 

Under oxidizing atmospheres cerium incorporates mainly as Ce$^{4+}$, which is not useful for the intended application as light emitter. Besides, partially fluctuations of the melt zone occurred.

\begin{figure}[htb]
\centering
\includegraphics[width=0.6\textwidth]{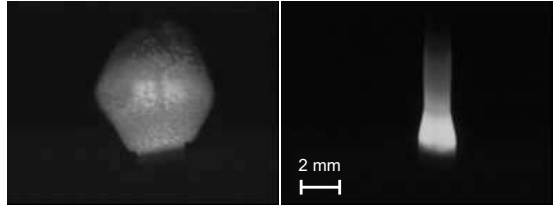}
\caption{Optical images taken during the growth process (left: SCF crystallizing in reducing atmosphere II; right: SCF crystallizing in inert atmosphere).}
\label{fig:dend}
\end{figure}

Under strongly reducing atmospheres II and III (Tab.~\ref{tab:growth_parameters}) cerium forms almost exclusively Ce$^{3+}$, but the process became less controllable and strong evaporation was observed. Often evaporating species condensed near the melt zone, forming a fine dendritic  felt shielding partially the melt zone from laser irradiation (Fig.~\ref{fig:dend}, left). Chemical analysis by ICP-OES indicated that calcium is the main constituent of this felt. Phase analysis was impossible due to its instability and minor quantity.

From Tab.~\ref{tab:atmosphere} it can be read that under strongly reducing conditions metallic Ca is the main evaporating component. Further calculations showed that the evaporating species react with other components of the atmosphere (mainly the rest oxygen $p_{\mathrm{O}_2}\approx2\times10^{-6}$\,bar) under the formation of solid CaO ($>98$\%) and small amounts of Ce$_2$O$_3$. One can assume that the dendrites consist of these components, and under the influence of water and carbon dioxide from ambient air CaO disintegrates quickly to Ca(OH)$_2$ and CaCO$_3$.

The best quality of SCFs was achieved by pulling in 99.999\% pure nitrogen at 1.8\,mm/min for the seed and 0.6\,mm/min for the feed rod (Fig.~\ref{fig:dend}, right). Then a usuful compromise between growth stability and useful Ce$^{3+}$ concentration was reached.

\begin{figure}[htb]
\centering
\includegraphics[width=0.6\textwidth]{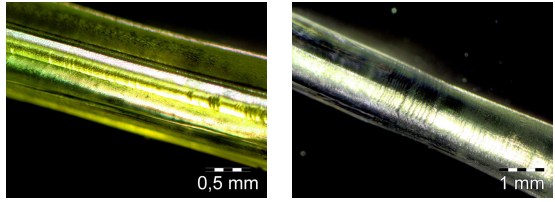}
\caption{LHPG grown calcium scandate SCFs (left: CaSc$_2$O$_4$:Ce$^{3+}$; right: CaSc$_2$O$_4$).}
\label{fig:crystals}
\end{figure}

Fig.~\ref{fig:crystals} shows optical images of doped and undoped SCFs. The undoped fiber is colorless, and the cerium doped fiber is light green to yellow green. Both show homogeneous extinction under cross-polarized light and are free of visible cracks. The surface of the SCFs is even. The undoped fiber shows periodic diameter variations. These variations evolved during crystallization from a melt that showed time-dependent convection. The origin of this convection is discussed in section~\ref{gas_phase}. When the convection of the melt was stationary, the fibers crystallized with planar surface (Fig.~\ref{fig:crystals} left). The maximum diameter of the LHPG fibers that could be reached was 1.6\,mm; at larger diameters the convection in the melt became time-dependent and the SCFs tended to crack parallel to the axis of symmetry.

\subsection{Phase characterization}
\label{xrd}

\begin{figure}[Htb]
\centering
\includegraphics[width=0.8\textwidth]{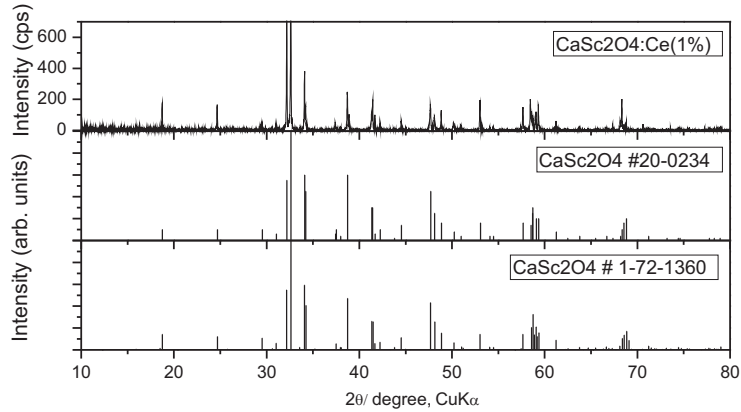} 
\caption{XRD patterns of a CaSc$_2$O$_4$:Ce SCF grown in argon (top panel) compared to 2 PDF files \cite{powder2011} of CaSc$_2$O$_4$.} 
\label{fig:crystal1}
\end{figure}

\begin{figure}[Htb]
\centering
\includegraphics[width=0.8\textwidth]{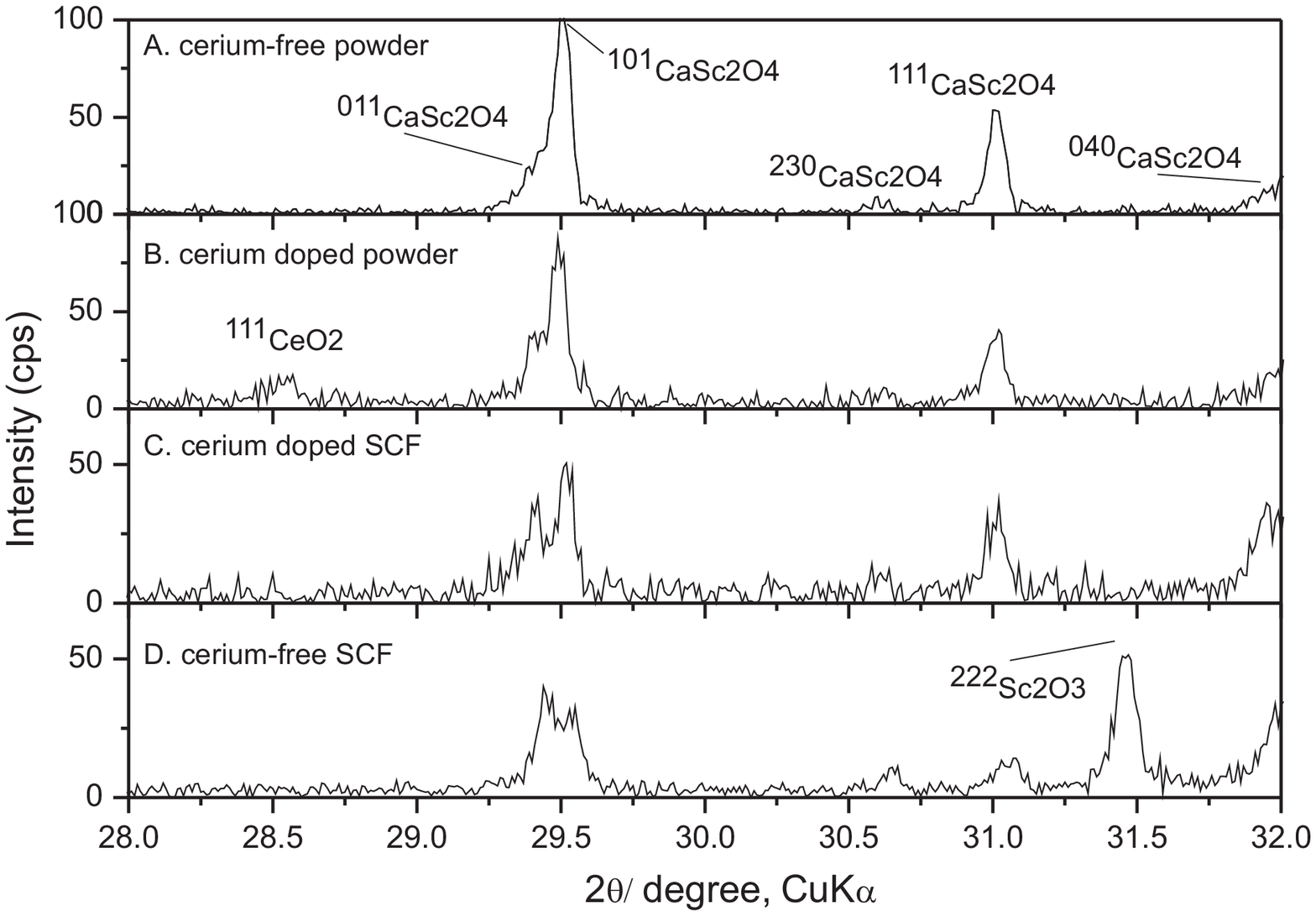} 
\caption{Powder XRD patterns of a cerium-free powder [A], cerium doped powder [B], cerium doped SCF [C] and cerium-free SCF [D] of CaSc$_2$O$_4$ and CaSc$_2$O$_4$:Ce(1\%). The powders were sintered at $1600^{\,\circ}$C for 24\,h in air; the SCFs were crystallized in inert atmosphere.} 
\label{fig:crystal2}
\end{figure}

Fig.~\ref{fig:crystal1} compares the powder XRD patterns of a CaSc$_2$O$_4$:Ce SCF with literature data. Neither peaks referring to CaO and its hydroxides, to Sc$_2$O$_3$ and cerium oxides, nor to other foreign oxide phases were found. All peaks except one at $2\theta=38.8^\circ$ can be indexed and referred to calcium scandate with the Powder Diffraction Files \cite{powder2011}. As this peak occurs in all XRD patterns of CaSc$_2$O$_4$, it might be associated with an intermetallic ScPt phase (PDF No. 00-019-0910) as a result of the high sintering temperatures in Pt crucibles, or it could be a relict of the K$_{\alpha2}$ 131 peak of CaSc$_2$O$_4$ (PDF No. 00-020-0234).

Fig.~\ref{fig:crystal2} gives a closer insight into the presence of impurities. Among these four powder XRD patterns only the cerium doped powder [B] shows a peak at $2\theta=28.68^\circ$. This is the strongest peak of CeO$_2$ (PDF No. 00-001-0800 \cite{powder2011}) that did not react completely after sintering at $1600^{\,\circ}$C in air. Nonetheless, the cerium doped SCF that was grown in argon atmosphere does not show this peak. The patterns [D] of the cerium-free SCF shows a peak at $31.44^\circ$. This is the 222 peak of Sc$_2$O$_3$. During LHPG growth of this fiber time-dependent convection of the melt occurred, in accordance with evaporation of CaO (see section~\ref{gas_phase}). Therefore, the existence of scandium oxide results from a shift of composition and could be related to the oscillation of the melt.

The XRD patterns in Fig.~\ref{fig:crystal2} demonstrate that LHPG is suitable for the growth of Ce doped CaSc$_2$O$_4$ fibers, without charge compensation by alkali ions, which is necessary, if phosphors are produced by the ceramic route \cite{SHIMOMURA2007}. The doping without charge compensation is probably related to the tunnel structure of CaSc$_2$O$_4$, where the Ca$^{2+}$ ions (112\,pm, eightfold coordinated) \cite{shannon1976} occupy the tunnels \cite{MUELLERBUSCHBAUM2003}. Ce$^{3+}$ ions (114\,pm, when eightfold coordinated), have nearly the same radius. Moreover, the CaFe$_2$O$_4$ structure type is described as a loose packed structure with several possible coordination numbers for the earth alkali ion site \cite{MUELLERBUSCHBAUM1963}. 

\subsection{Gas phase and melt zone studies}
\label{gas_phase}

\begin{figure}[htb]
\centering
\includegraphics[width=0.8\textwidth]{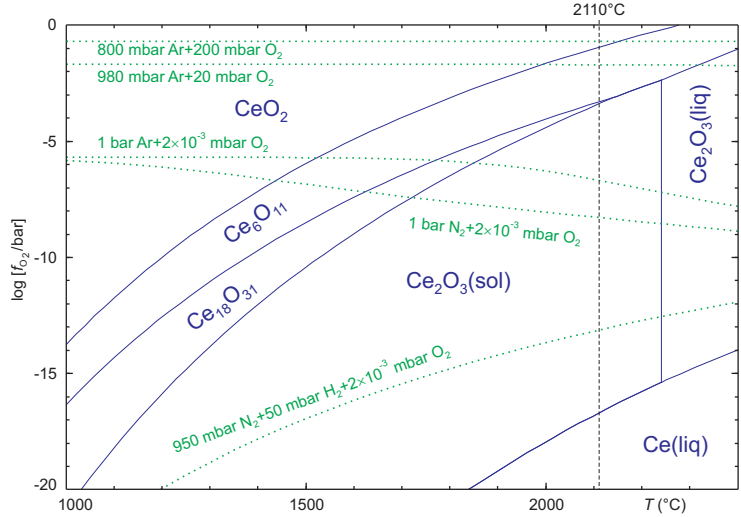} 
\caption{Predominance diagram Ce--O$_2$ (blue solid lines) showing the stability regions of Ce and its oxides as functions of oxygen fugacity $f_{\mathrm{O}_2}$ and temperature $T$. The green dotted lines show the $f_{\mathrm{O}_{2}}(T)$ that are supplied by different atmospheres used during LHPG growth. The dashed vertical line indicates the melting point of calcium scandate $T_\mathrm{f}=2110^{\,\circ}$C.}
\label{fig:atmo}
\end{figure}

The composition of the atmosphere affect both the evaporation of species and the oxidation number of cerium. The latter is demonstrated in Fig.~\ref{fig:atmo} that was calculated with Factsage\texttrademark. It should be repeated here that no atmosphere is really free of oxygen, because even for a technical gas with 99.999\% purity the rest gas supplies a background of typical $f_{\mathrm{O}_2}\approx2\times10^{-6}$\,bar (see Tab.~\ref{tab:growth_parameters}).

For the incorporation of trivalent cerium into the fiber the oxygen fugacity must be located within the stability region of Ce$_2$O$_3$. This region has in Fig.~\ref{fig:atmo} an upper limit $<10^{-20}$\,bar at $1000^{\,\circ}$C and extends to almost 0.1\,bar at the maximum $T=2400^{\,\circ}$C shown there. If $f_{\mathrm{O}_{2}}(T)$ exceeds these limits, the higher oxides Ce$_{18}$O$_{31}$, Ce$_6$O$_{11}$, or even CeO$_2$ are formed which contain less or no Ce$^{3+}$. It is desirable, to remain in the Ce$^{3+}$ phase fields not only at the growth temperature and slightly above (because the melt zone is overheated), but also at lower $T$, because even if Ce$^{3+}$ is stable at $T_\mathrm{f}$, it might be subsequently oxidized at lower $T$, when the grown fiber is removed from the melt zone.

The reducing atmosphere meets this condition over a wide temperature range, and consequently fiber growth in this atmosphere is preferable with respect to the Ce$^{3+}$ content. Unfortunately, the formation of CaO dendrites close to the fiber crystallization front impedes the application of strongly reducing atmospheres containing H$_2$ (Fig.~\ref{fig:dend} left). So-called ``inert'' 5N Ar or 5N N$_2$ are from the chemical viewpoint less advantageous, but it is obvious from Fig.~\ref{fig:atmo} that they represent a compromise that holds at least down to several 100\,K below $T_\mathrm{f}$. Even if an identical rest gas impurity $p_{\mathrm{O}_2}=2\times10^{-6}$\,bar is assumed for both gases, nitrogen is superior over argon, because at high $T$ $f_{\mathrm{O}_2}$ is reduced there due to the formation of nitrogen oxides, mainly NO. (The less significant drop of $f_{\mathrm{O}_2}(T)$ also for Ar results from the partial dissociation of O$_2$.) Experimentally it could be shown that SCFs grown in ``inert'' or slightly reducing atmosphere, especially in 5N nitrogen, contained a significant amount of Ce$^{3+}$ \cite{Physik2011}.

Oxidizing atmospheres do not cross the stability region of Ce$_2$O$_3$ (Fig.~\ref{fig:atmo}), but reduce the evaporation especially of calcium. This is shown in Tab.~\ref{tab:atmosphere} for the relevant species Ca, CaO, Sc, ScO, CeO and CaOH. Several restrictions of these equilibrium calculations have to be acknowledged, however: 1. Only a small volume of the system is at the estimated melt zone temperature of $2250^{\,\circ}$C. 2. Within the melt zone there is a high temperature gradient. 3. Due to the high growth rates, the system might not be in perfect equilibrium.

\begin{table}[htbp]
\centering
\begin{tabular}{lllllll}
\hline
Atmosphere (mbar)        &                              \multicolumn{6}{c}{Fugacities (bar)} \\
\hline
                         & Ca               & CaO                  & Sc                 & ScO               & CeO              & CaOH  \\
\hline
Ar (980), O$_2$ (20)     & $4.6\times10^{-6}$ & $9.3\times10^{-6}$ & -                  & $1.2\times10^{-7}$& $7.5\times10^{-7}$ & - \\
Ar (1000)                & $1.2\times10^{-4}$ & $9.3\times10^{-6}$ & -                  & $6.1\times10^{-7}$& $4.8\times10^{-6}$ & - \\
N$_2$ (1000)             & $2.2\times10^{-4}$ & $9.3\times10^{-6}$ & -                  & $8.1\times10^{-7}$& $6.4\times10^{-6}$ & - \\
N$_2$ (990), H$_2$ (10)  & $9.0\times10^{-4}$ & $9.3\times10^{-6}$ & -                  & $1.6\times10^{-6}$& $1.0\times10^{-5}$ &$6.8\times10^{-5}$ \\
N$_2$ (950), H$_2$ (50)  & $9.5\times10^{-4}$ & $1.9\times10^{-6}$ & $4.8\times10^{-8}$ & $3.7\times10^{-6}$& $9.9\times10^{-6}$ &$3.4\times10^{-5}$ \\
\hline 
\end{tabular}
\caption{Fugacities $f$ of species evaporating from the melt at $2250^{\,\circ}$C in different atmospheres. A background oxygen fugacity of $f_{\mathrm{O}_2}=2\times10^{-6}$\,bar was added always.}
\label{tab:atmosphere}
\end{table} 

For all atmospheres, species containing Ca are dominantly evaporating, whereas evaporation of Sc and Ce species is subordinated. For inert gases that gave the best fibers, the fugacity of CeO is almost two orders of magnitude smaller than Ca, but the initial Ce content within the fiber is also smaller by two orders of magnitude. Therefore, a comparable deficit of Ca and Ce within the fibers can be expected.

\begin{figure}[htb]
\centering
\includegraphics[width=0.7\textwidth]{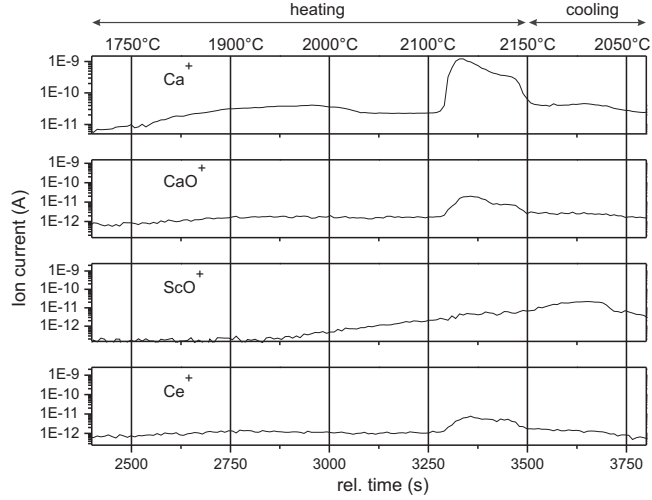} 
\caption{Monitored ion intensities of Ca$^{+}$,CaO$^{+}$, ScO$^{+}$ and Ce$^{+}$ during heating of a CaSc$_2$O$_4$:Ce SCF up to $2150^{\,\circ}$C. The temperature was adjusted manually.}
\label{fig:htms}
\end{figure}

The significance of the calculated fugacities that are shown in Tab.~\ref{tab:atmosphere} was checked by measurements with a high temperature mass spectrometer (Fig.~\ref{fig:christo}). Fig.~\ref{fig:htms} shows ion intensities of relevant species during heating of a CaSc$_2$O$_4$:Ce SCF. Other evaporating species were monitored but not shown here, because their signals were low. Major evaporation of Ca starts at approximately $1800^{\,\circ}$C. This implies, the composition might be shifted already during the repeated sintering of the feed rods at $1600^{\,\circ}$C. The ScO$^+$ signal increases eventually after evaporation of Ca at $\approx2030^{\,\circ}$C. The formation of gaseous ScO is due to partial melt of the fiber surface that is depleted of Ca. The melting point at the fiber surface is lowered and seems to lie at $\approx2030^{\,\circ}$C, accompanied by a decrease of the Ca$^+$ signal. High amounts of Ca, CaO and Ce are evaporating when the bulk of the fiber melts near $2110^{\,\circ}$C. These results of the kinetic measurements are in agreement with the predicted calculations under equilibrium conditions and the DTA results (section~\ref{thermal}). With reference to the MF-HTMS results it seems reasonable that Ce$^{3+}$ is stabilized by the formation of Ca vacancies as it is suggested for the CaSc$_{2}$O$_{4}$:Ce powder \cite{SHIMOMURA2007, CHEN2011}.

\begin{table}[htb]
\centering
\begin{tabular}{llllr}
\hline
\multicolumn{1}{c}{Sample}            & \multicolumn{3}{c}{Rel. deviation (\%)} & \multicolumn{1}{r}{Sample Number}  \\ 
\hline 
                                      &  Ca    &   Sc               &  Ce       & \\
\hline
Feed rod                              & -0.9   & +0.5               & -5.4      & CS2-P3 \\

SCF (1000\,mbar N$_2$)                & -4.4   & +2.5               & -56.1     & CS2-52 begin\\
SCF (1000\,mbar N$_2$)                & -4.3   & +2.3               & -40.9     & CS2-52 end\\
FZ (1000\,mbar N$_2$)                	& -4.4   & +2.1               & -11.6     & CS2-52 frozen melt zone\\

SCF (990\,mbar N$_2$, 10\,mbar H$_2$) & -6.2   & +3.4               & -68.9     & CS2-53 begin \\
SCF (990\,mbar N$_2$, 10\,mbar H$_2$) & -11.2  & +5.5               & -14.5     & CS2-53 mid \\
SCF (990\,mbar N$_2$, 10\,mbar H$_2$) & -5.6   & +2.9               & -13.3     & CS2-53 end \\
FZ (990\,mbar N$_2$, 10\,mbar H$_2$)  & -14.1  & +6.2               & +146.7    & CS2-53 frozen melt zone \\

SCF (950\,mbar N$_2$, 50\,mbar H$_2$) & -22.4  & +11.4              & -63.1     & CS2-51 begin\\
SCF (950\,mbar N$_2$, 50\,mbar H$_2$) & -20.8  & +10.5              & -37.9     & CS2-51 end\\
FZ 	(950\,mbar N$_2$, 50\,mbar H$_2$) & -20.2  & +9.6             	& +66.1     & CS2-51 frozen melt zone\\

\hline
\end{tabular}
\caption{Relative deviation (from the weighed portion) of the elements within an air sintered feed rod of CaSc$_{2}$O$_{4}$:Ce, SCFs and their frozen melt zones crystallized in different atmospheres: CS2-52 (1000\,mbar N$_2$), CS2-53 (990\,mbar N$_2$, 10\,mbar H$_2$) and CS2-51 (950\,mbar N$_2$, 50\,mbar H$_2$) (elementary analysis with ICP-OES).}
\label{tab:rel_deviation}
\end{table}

Elementary analysis (ICP-OES) of LHPG fibers gave further information on the composition shift. Tab.~\ref{tab:rel_deviation} shows the relative deviation of the components for an air sintered feed rod, SCFs and their frozen melt zones that crystallized in different atmospheres: CS2-52 (1000\,mbar N$_2$), CS2-53 (990\,mbar N$_2$, 10\,mbar H$_2$) and CS2-51 (950\,mbar N$_2$, 50\,mbar H$_2$). The deviations can be explained by the calculated fugacities (Tab.~\ref{tab:atmosphere}) and experimental MF-HTMS results (Fig.~\ref{fig:htms}). The feed rod that was sintered at $1600^{\,\circ}$C has a negative deviation of Ca and Ce, whereas Sc is enriched. The fibers show a greater loss of calcium, when crystallized in atmospheres containing more hydrogen: -4.4\% calcium deviation in the SCF CS2-52 (no hydrogen), -6.2\% in the SCF CS2-53 (1\% hydrogen) and -22.4\% in SCF CS2-51 (5\% hydrogen). Some of the cerium evaporated during feed rod preparation (-5.4\%). SCF CS2-53 indicates the steady state condition of the LHPG process: Between the middle part, approximately 4 mm above the first crystallized piece of the SCF, and the end of the fiber only small changes of the deviation occur. Taking account of the mass lost due to the evaporation, steady state conditions are reached and a constant mass flow from the feed to the melt zone and from the melt zone into the fiber is attained. 

The effective distribution coefficient $k_\mathrm{eff}$ can be estimated by 
\begin{equation}
k_\mathrm{eff} \approx \frac{C_\mathrm{i}}{C_0} \approx \frac{C_\mathrm{e}}{C_\mathrm{fz}}
\end{equation}
with $C_\mathrm{i}$ and $C_\mathrm{e}$ as dopant concentration at the begin and the end of the fiber and $C_\mathrm{fz}$ as dopant concentration in the frozen melt zone \cite{Pfann1958}. We obtained an average $k_\mathrm{eff}=0.4$ for the reducing atmosphere III (950\,mbar N$_2$, 50\,mbar H$_2$) respectively $k_\mathrm{eff}=0.5$ for the reducing atmosphere I (1000\,mbar N$_2$) and pure argon atmosphere.

During the LHPG growth time-dependent Marangoni convection of the melt zone was observed. Among other high melting oxides crystallized with LHPG by the present authors, Sc$_2$O$_3$:Ce, Lu$_2$O$_3$, SrY$_2$O$_4$:Ce, Sr$_3$SiO$_5$:Ce, only calcium scandate showed this oscillating convection. The observed frequency of 1\,Hz agrees with the nonsteady thermocapillar convection described by Jurisch and Loeser \cite{jurisch1990}. We could minimize the oscillation by reducing the fiber diameter and increasing the growth rate. Several effects resulted in these changes: 1. reduction of evaporation due to a higher mass flow; 2. reduction of the radial temperature gradient according to Prokofiev et al. \cite{prokofiev1995}. According to Schwabe and Scharmann \cite{SCHWABE1979} oscillatory thermocapillary convection (OTC) results from exceeding some critical Marangoni number. This transgression could be attained during LHPG growth by a shift of composition and a subsequent overheating of the melt.

\subsection{Photoluminescence}
\label{photo}

\begin{figure}[htb]
\centering
\includegraphics[width=0.7\textwidth]{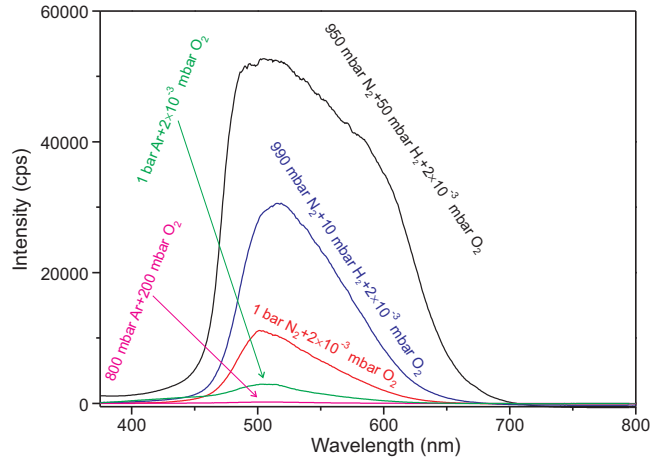} 
\caption{Photoluminescence spectra of SCFs that were crystallized in different atmospheres. The photoluminescence excitation was at 337\,nm.} 
\label{fig:lumin}
\end{figure}

If the photoluminescence spectra are taken into account, the relation between the influence of the atmosphere and the resulting content of trivalent cerium becomes vivid. Fig.~\ref{fig:lumin} shows the photoluminescence (excitation at 337\,nm) of SCFs, grown in the atmospheres that were already discussed regarding the fugacity and deviation from the weighed portion in \ref{gas_phase}. Especially the black luminescence curve of a fiber, crystallized in the reducing atmosphere with 5\% hydrogen, has the same shape as the curve of the CaSc$_{2}$O$_{4}$:Ce phosphor that has been investigated by Shimomura \cite{SHIMOMURA2007}. The photoluminescence originates from the $5d-4f$ transition of trivalent cerium. The broadened structure with its maximum at 515\,nm is due to the ground state spliting of the Ce$^{3+}$ ion \cite{SHIMOMURA2007}. Whereas the total amount of cerium within the fiber does not change significantly (ICP-OES), the amount of trivalent cerium is strongly influenced by the oxygen fugacities of the atmospheres (as it is also shown by the Factsage\texttrademark~calculations in \ref{gas_phase}). The photoluminescence is lowered with higher values of $f_{\mathrm{O}_2}$ and cannot be detected in a SCF that crystallized in an oxidizing atmosphere (purple curve in Fig.~\ref{fig:lumin}). Crystallizing in 5N nitrogen atmosphere causes a significant luminescence (red curve), but a certain amount of Ce$^{4+}$ must be taken into account.

\subsection{Thermal Analysis}
\label{thermal}

\begin{figure}[htb]
\centering
\includegraphics[width=0.6\textwidth]{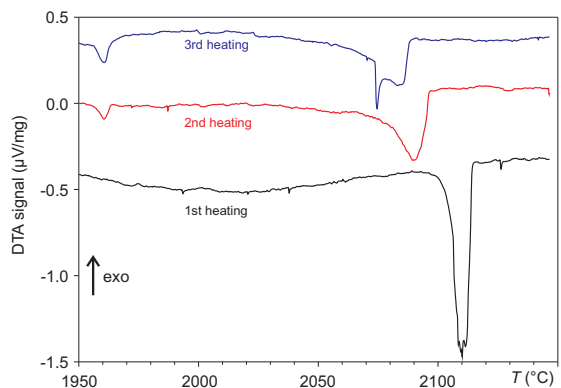} 
\caption{Subsequent DTA heating curves of a CaSc$_2$O$_4$ fiber.}
\label{fig:dta1}
\end{figure}

Fig.~\ref{fig:dta1} shows three subsequent DTA heating curves of an undoped CaSc$_2$O$_4$ fiber with identical heating rates of 15\,K/min. Cooling curves were also recorded but not shown here, because the crystallization peaks appearing there showed some remarkable (several 10\,K) but badly reproducible supercooling with respect to the heating peaks. Thus, these curves are so far not contributing to a better understanding of phase relations.

The first heating curve shows the melting point with onset at $2110^{\,\circ}$C, in total agreement with the value given by Get'man \cite{GETMAN2000}. During the second and third heatings the peak shifts to lower $T$ and its area becomes smaller. Besides, an additional peak with an onset at $1958^{\,\circ}$C appears. It was found that mainly Ca evaporates from the sample, consequently the composition shifts to the Sc-rich side of the system CaO--Sc$_2$O$_3$. Because no other pseudobinary Ca-Sc oxide phases except CaSc$_2$O$_4$ were found in the system (section~\ref{xrd}), this additional peak is assumed to mark the eutectic point between Sc$_2$O$_3$ and CaSc$_2$O$_4$. 

\section{Conclusions}
\label{conc}

First single crystal fibers of Ce$^{3+}$ doped CaSc$_2$O$_4$ have been grown using laser-heated pedestal growth. The establishment of optimum growth conditions required a compromise between the stabilization of Ce$^{3+}$ instead of Ce$^{4+}$ (which is possible in reducing growth atmosphere) and chemical stability of the CaSc$_2$O$_4$ melt with respect to Ca evaporation (which is possible in oxidizing growth atmosphere). The rest oxygen impurity content of technical nitrogen gas with 5N (99.999\%) nominal purity proved to deliver appropriate conditions. Besides, large growth rates should be applied to keep Ca and Ce loss due to evaporation small.

Composition changes of the melt, resulting from evaporation of its constituents, can be described reasonably by thermodynamic equilibrium calculations. The significance of these calculations was demonstrated by chemical analysis of the solids with ICP-OES, by evolving gas analysis with mass spectrometry during heating, and by thermal analysis up to the melting point $T_\mathrm{f}=2110^{\,\circ}$C of CaSc$_2$O$_4$.

For an initial composition Ca$_{0.99}$Ce$_{0.01}$Sc$_2$O$_{4.01}$ a distribution coefficient $k_\mathrm{eff}\approx0.5$ was measured for the cerium dopant in 5N nitrogen atmosphere. Spectroscopic measurements showed that a significant part of this cerium is trivalent, and its application as a new laser material \cite{Physik2011} seems feasible.

\section{Acknowledgment}
\label{ackno}
The authors are indebted to K. Irmscher and A. Kwasniewski  (both Leibniz Institute for Crystal Growth Berlin) for photoluminescence and X-ray powder diffraction measurements. Fruitful discussions with S. Ganschow and R. Uecker are highly acknowledged. 









\end{document}